\documentclass[prd, twocolumn, nofootinbib, floatfix]{revtex4-1}
\pdfoutput=1

\usepackage{hyperref}
\usepackage{amsmath}
\usepackage{graphicx}
\usepackage{dcolumn}
\usepackage{bm}
\usepackage{epsfig}
\usepackage{amssymb,latexsym,mathrsfs}
\usepackage{graphicx}
\usepackage{color}

\hypersetup{
    colorlinks=true,
    linkcolor=red,
    citecolor=blue,
    urlcolor=magenta,
} 

\newcommand{\be}{\begin{equation}}
\newcommand{\ee}{\end{equation}}
\newcommand{\bea}{\begin{eqnarray}}
\newcommand{\eea}{\end{eqnarray}}

\newcommand{\om}{\Omega_m} 
 
\newcommand{\dlz}{\delta z} 
\newcommand{\dchi}{\Delta\chi^2}

\begin{document}

\title{Photometric Supernovae Redshift Systematics Requirements} 

\author{Eric V.\ Linder${}^{1,2}$, Ayan Mitra${}^2$} 
\affiliation{
${}^1$Berkeley Center for Cosmological Physics \& Berkeley Lab, 
University of California, Berkeley, CA 94720, USA\\ 
${}^2$Energetic Cosmos Laboratory, Nazarbayev University, 
Nur-Sultan, Kazakhstan 010000 
}

\date{\today}

\begin{abstract}
Imaging surveys will find many tens to hundreds of thousands of Type Ia 
supernovae in the next decade, and measure their light curves. In addition 
to a need for characterizing their types and subtypes, a redshift is required 
to place them on a Hubble diagram to map the cosmological expansion. We 
investigate the requirements on redshift systematics control in order not 
to bias cosmological results, in particular dark energy parameter estimation. 
We find that additive and multiplicative systematics must be constrained at 
the few$\,\times 10^{-3}$ level, effectively requiring spectroscopic followup for robust 
use of photometric supernovae. Catastrophic outliers need control at the 
subpercent level. We also investigate sculpting the spectroscopic sample. 
\end{abstract} 

\maketitle

%%%%%%%%%%%%%%%%%%%%%%%%%%%%%%%%%%%%%%%%%%%%%%%%%%%%%%%%%%%% 
\section{Introduction}

Type Ia supernovae (SN Ia) are standardizable distance measures, whose use 
led to the discovery of cosmic acceleration \cite{perl99,riess98}, and still 
provide the most stringent constraints on the nature of dark energy 
\cite{review1,review2,union2p1,jla,pantheon}. In the next decade, the number of 
SN Ia discovered and imaged in multiple photometric wavelength bands will 
increase by a factor of $\sim100$, driven by surveys such as the Zwicky 
Transient Factory (ZTF \cite{ztf}) and the Large Synoptic Survey Telescope 
(LSST \cite{lsst}). If those SN can be fully utilized for 
cosmology they will provide powerful leverage on uncovering the nature of 
cosmic acceleration. 

However, even with current samples systematics contribute at least equally 
to statistical uncertainty in the cosmological use of SN Ia. These 
systematics can be addressed through careful characterization of the 
supernova properties, through enhanced wavelength coverage into the 
infrared \cite{snir,snir1} and ultraviolet \cite{snuv,snuv1}, and in particular 
spectroscopic data \cite{snspec,snspec1,snspec2}. Spectroscopy not only confirms the source 
to be a true SN Ia but also give subtyping, e.g.\ through line ratios, high 
vs low velocities, etc. This then permits matching of similar SN Ia at 
different redshifts, greatly ameliorating systematics, through ``like vs 
like'' \cite{likes,likes2,0812.0370} or more detailed ``twinning'' \cite{hannah,hannah2} 
methods. 

The next decade imaging surveys (and the recently completed Dark Energy 
Survey \cite{des}) rely on spectroscopic follow up to obtain the detailed 
information, as well as accurate redshifts. This limits the most robust 
sample to a few hundred in the case of Dark Energy Survey or a few thousand 
for next decade surveys, due to the time requirements for making the 
spectroscopic measurements. While there will be next decade multiobject 
spectroscopic instruments such as the Dark Energy Spectroscopic Instrument 
(DESI \cite{desi,desi2}) and 4-metre MultiObject Spectroscopic 
Telescope (4MOST \cite{4most}), the relatively low 
multiplexing of SN Ia observations means that many will not have spectroscopic 
data. 

This has led to an extensive literature exploring whether purely photometric 
measurements can robustly place SN Ia on the Hubble diagram (see, e.g., 
\cite{syst1,syst2,syst3,syst4,syst5,syst6,syst7,syst7b,syst8,syst9,syst10,syst11} among others). Issues include contamination by 
non-SN Ia, lack of subtyping, and selection effects. These all can distort 
the Hubble diagram vertically, by misestimating the source distance. Here 
we focus on redshift errors, which biases the Hubble diagram horizontally. 

Note that a common practice is to obtain the source redshift by measuring 
the spectroscopic redshift of the host galaxy. If successful, this is 
adequate, but such a measurement still requires telescope time and becomes 
increasingly expensive at higher redshifts where leverage on dark energy may 
be greater. Galaxy catalog redshifts will also become more incomplete as one 
goes to the higher redshifts accessed by the next decade surveys. Moreover, 
not all SN Ia will have readily (or uniquely) identified hosts \cite{hosts,hosts1,hosts2}. 
Photometric redshifts give a rough indication of the source redshift, 
but face challenges in use for accurate cosmology. 

The redshift requirements for the Hubble diagram were investigated in 
the pioneering article of \cite{0402002} (and later \cite{0607030} for 
high redshift). They propagated redshift 
uncertainties into cosmological parameter uncertainties, i.e.\ how a finite 
prior on the mean redshift within a bin of supernovae nearly at the same 
distance increased the cosmology uncertainty. Their conclusion is that the 
redshift must be known to 0.002 or better to limit the increase in 
uncertainty on a constant dark energy equation of state to less than 10\%. 
Here we investigate the complementary issue of systematic bias -- shift in 
derived cosmology -- rather than increase in statistical uncertainty. 

In Sec.~\ref{sec:zbias} we present the redshift systematic and cosmological 
parameter bias formalism. We assess the impact, and derive the requirements 
on systematic control in Sec.~\ref{sec:results}, for additive, multiplicative, 
and catastrophic systematics. In Sec.~\ref{sec:concl} we discuss the 
results and conclude.

%%%%%%%%%%%%%%%%%%%%%%%%%%%%%%%%%%%%%%%%%% 
\section{Redshift Systematics and Cosmology Bias} \label{sec:zbias} 

When the measurement of an observable is systematically offset from its 
true value, the cosmology estimation following from the data will be biased. 
For small offsets, the Fisher bias formalism provides a straightforward 
technique for investigating the size and impact of this effect. The bias 
on a cosmological parameter is given by \cite{fisbias1,fisbias2} 
\be 
\delta p_i=\left(F^{-1}\right)_{ij}\sum_k \frac{\partial{\mathcal O_k}}{\partial p_j}\frac{1}{\sigma_k^2}\Delta{\mathcal O_k}\ , 
\ee 
where $F^{-1}$ is the inverse Fisher matrix, 
${\mathcal O_k}$ is the observable, $\sigma_k$ its 
uncertainty, and $\Delta{\mathcal O_k}$ is the offset in the observable due 
to the systematics. The expression here takes a simple diagonal noise matrix. 

For the SN Ia case, the observable is the apparent magnitude $m$ 
(really derived from observed photometry and redshift), and the 
offset is due to systematic misestimation of the redshift, so 
\be 
\Delta{\mathcal O}(z)=\frac{\partial m}{\partial z}\,\dlz\ . 
\ee 
The partial derivative involves two components, the change in the distance 
modulus or luminosity distance $d_L$ with redshift, and how the change 
in redshift alters the relation between the apparent magnitude and distance 
modulus. The latter part is due to the data analysis procedure for 
standardization based on light curve width and color or dust extinction. 

From \cite{0402002} we see that the wavelength dependent color ($k$-correction) 
and extinction factor varies rapidly and with spikes in the redshift as the 
rest frame SN flux moves through the various photometric survey bands. 
This does not resemble the smooth variation from a cosmological parameter 
and so will not significantly bias cosmology estimation 
\cite{0812.0370,0908.2637,kcrx1,kcrx2}. (We have verified this numerically with a 
spiking toy model.) The light curve width does contain a component 
from time dilation, proportional to $1+z$, and so the standardization 
procedure with an incorrect redshift will cause a shift in $m$. Thus we 
take 
\be 
\frac{\partial m}{\partial z}=\frac{\partial m}{\partial d_L}\, 
\frac{\partial d_L}{\partial z}+
\frac{\partial m}{\partial{\rm width}}\,\frac{\partial{\rm width}}{\partial z}
\ . 
\ee 

Since $m\sim 5\log[d_L(z)]$, the factor 
\be 
\frac{\partial m}{\partial d_L}=\frac{5}{\ln 10}\,\frac{1}{d_L(z)}\ . 
\ee 
Recall that $d_L(z)=(1+z)\int_0^z dz'/[H(z')/H_0]$, for a flat universe 
as we will assume, with $H$ the Hubble parameter. 
In common light curve width standardization methods, the apparent magnitude 
is linearly proportional to the rest frame light curve width, and so 
\be
\frac{\partial m}{\partial{\rm width}}={\rm const}\ . 
\ee 
For example, in stretch standardization $m\sim -\alpha(s-1)$ and in 
SALT2 \cite{salt2} light curve fitting $m\sim -\alpha' X_1$, where 
$\alpha$ and $\alpha'$ are constants. Since the rest frame width equals 
the observer frame width divided by $1+z$, then, e.g., 
$ds=-s\,dz/(1+z)$. The analogous expression holds for SALT2. 

Putting this all together we have 
\be
\Delta m=\frac{5}{\ln 10}\,\ln\left[\frac{d_L(z+\dlz)}{d_L(z)}\right] 
+\frac{C\,\dlz}{1+z}\ . 
\ee 
For $\dlz\ll z$ one could expand the distance ratio by evaluating the 
derivative $d\,d_L/dz$; we do not do this since for low redshift SN we 
may not have $\dlz\ll z$, and in any case the derivative would still leave 
integrals to be evaluated and so does not save much effort. However we 
can note that we expect $\Delta m \propto \dlz$ to a good approximation. 
The constant $C\approx\alpha s\approx\alpha' (X_1+1)\approx 1.4$ 
averaging over supernovae \cite{salt2}. 

The cosmological parameters $p_i$ are the matter density $\Omega_m$ in 
units of the critical density, the dark energy equation of state parameter 
today $w_0$ and a measure of its time variation $w_a$, and the combination 
${\mathcal M}$ of the SN absolute magnitude and Hubble constant. 
When quoting constraints on one parameter we marginalize over 
the other parameters. 

We now have to specify the survey properties, i.e.\ the number and 
distribution of SN Ia and their magnitude uncertainty. We do not attempt 
to model a next decade survey, with all its real world selection effects; 
rather we adopt a simple model that should be a reasonable approximation. 
To a statistical dispersion of $\sigma_{\rm stat}=0.15$\,mag per SN Ia 
(reasonable for a photometric survey), we add in quadrature a systematic measurement 
floor of $\sigma_{\rm sys}=0.01\,(1+z)$ per redshift bin of width 0.1. 
That is, infinite numbers of SN Ia will not give infinite accuracy, but 
rather uncertainties will be limited by the floor, representing, e.g., 
photometric band calibration zeropoint uncertainties, light curve model 
uncertainties, survey selection effects, etc. 

Table~\ref{tab:floor} 
summarizes the number of SN Ia used in each redshift bin, and the ratio of 
the systematic to the total magnitude error, showing that more 
statistics will not help significantly. (And indeed if the error is 
lowered, any bias will become more severe in a relative sense.) We take a 
redshift range of $z=0-1.2$ (though the $z<0.1$ SN may come from a separate 
survey). One can regard this as a reasonable, if rough, approximation to 
a next decade SN survey.

%%%%%%%%%%%%%%%%%%%%%%%%%%% 
\begin{table}[htbp!]
\begin{tabular}{|c|c|c|c|}
 \hline
$z$ & $n$ & $\sigma_{\rm tot}$ & \ sys/total\ \\ 
\hline 
\ 0.05 \ & \ 300 \  & \ 0.014 \  & 0.77 \\ 
0.15 & 300 & 0.014 & 0.80 \\ 
0.25 & 300 & 0.015 & 0.82 \\ 
0.35 & 300 & 0.016 & 0.84 \\ 
0.45 & 300 & 0.017 & 0.86 \\ 
0.55 & 300 & 0.018 & 0.87 \\ 
0.65 & 300 & 0.019 & 0.89 \\ 
0.75 & 300 & 0.020 & 0.90 \\ 
0.85 & 300 & 0.020 & 0.91 \\ 
0.95 & 300 & 0.021 & 0.91 \\ 
1.05 & 150 & 0.024 & 0.86 \\ 
1.15 & 150 & 0.025 & 0.87 \\ 
\hline
 \end{tabular}  
\caption{
Survey characteristics adopted as an approximation of a next decade 
survey in terms of total magnitude uncertainty $\sigma_{\rm tot}$. This 
is given in magnitudes and is the most important property. The last column 
shows that the total uncertainty is predominantly systematics dominated, so 
the number $n$ of SN Ia in each redshift $z$ bin is mostly moot. 
 } 
\label{tab:floor}
\end{table}

Finally, we need to specify a model for $\dlz$. We take it to be systematic among all supernovae at redshift $z$ and consider three types of 
redshift systematics: additive, multiplicative, and catastrophic errors. 
The first two are simply described by 
\be 
\dlz=d_0+d_1z\ .
\label{e1}
\ee 
As long as $\dlz\ll z$ one can show that the parameter biases $\delta p_i$ 
are linear in $\dlz$ and so one can explore the impact of $d_0$ (additive) and $d_1$ (multiplicative) 
separately. One can then add the $\delta p_i$ afterward if desired, or 
take the individual effects as a lower limit on the systematics control 
required. 

Catastrophic redshift errors are more sensitive to the complicated 
survey characteristics so we only adopt three toy models to give a rough 
estimation of their effects. For each, we assume a fraction $f$ of the 
SN Ia in each redshift bin are affected, and derive the control needed 
on $f$. The first model puts misestimated SN from each redshift at $z=0.1$, and the 
second model puts them at $z=1$, regardless of their true redshift. The 
third model puts SN from true redshifts $z<0.6$ at $z+0.2$ and SN from 
true redshifts $z>0.6$ at $z-0.2$. That is, it narrows the redshift 
distribution. 

Given all the elements we propagate the redshift systematics into 
cosmological parameter biases. Since the covariance between parameter 
shifts is important -- i.e.\ a modest shift orthogonal to the degeneracy 
direction can place the derived values well outside the true joint confidence 
contour -- we quantify this by evaluating the change in likelihood due 
to the bias \cite{shapiro,shapiro2}, 
\be 
\Delta\chi^2={\bm{\delta p}}\,{\bm F}^{\rm sub}\,({\bm{\delta p}})^T\ , 
\ee 
where we define a subspace of interest, e.g.\ the dark energy $w_0$--$w_a$ 
plane, and convolve the Fisher submatrix (marginalized over other parameters) 
with the parameter bias vectors. In all calculations 
we include a Planck prior on the distance to CMB last scattering.

%%%%%%%%%%%%%%%%%%%%%%%%%%%%%%%%%%%%%%%%%% 
\section{Systematics Requirements} \label{sec:results} 

For each of the forms of the redshift systematics we calculate the 
cosmological parameter biases, and the $\dchi$ in the $w_0$--$w_a$ plane, 
i.e.\ the offset of dark energy properties relative to the true joint 
likelihood (marginalized over the other parameters). Due to the linearity 
of $\delta p_i$ with respect to $\dlz$ we can estimate the systematics 
control necessary, in terms of limiting $d_0$, $d_1$, or $f$, so as to 
ensure bias is not significant. Note that a condition such as, say, $\delta p_i<\sigma(p_i)/2$ is not sufficient to 
prevent a large shift in terms of $\dchi$, since that involves a nonlinear combination of various 
parameters $p_i$ and their covariances. Therefore we evaluate $\dchi$ and impose  $\dchi<2.30$, 
i.e.\ limiting the misestimation of the dark energy properties to stay 
within the $1\sigma$ joint confidence contour of $w_0$--$w_a$. 
This gives the requirements on the systematics control.

%%%%%%%%%%%%%%%%%%% 
\subsection{Additive Systematic} \label{sec:add} 

\begin{figure}[htbp!] 
\centering
\includegraphics[width=\columnwidth]{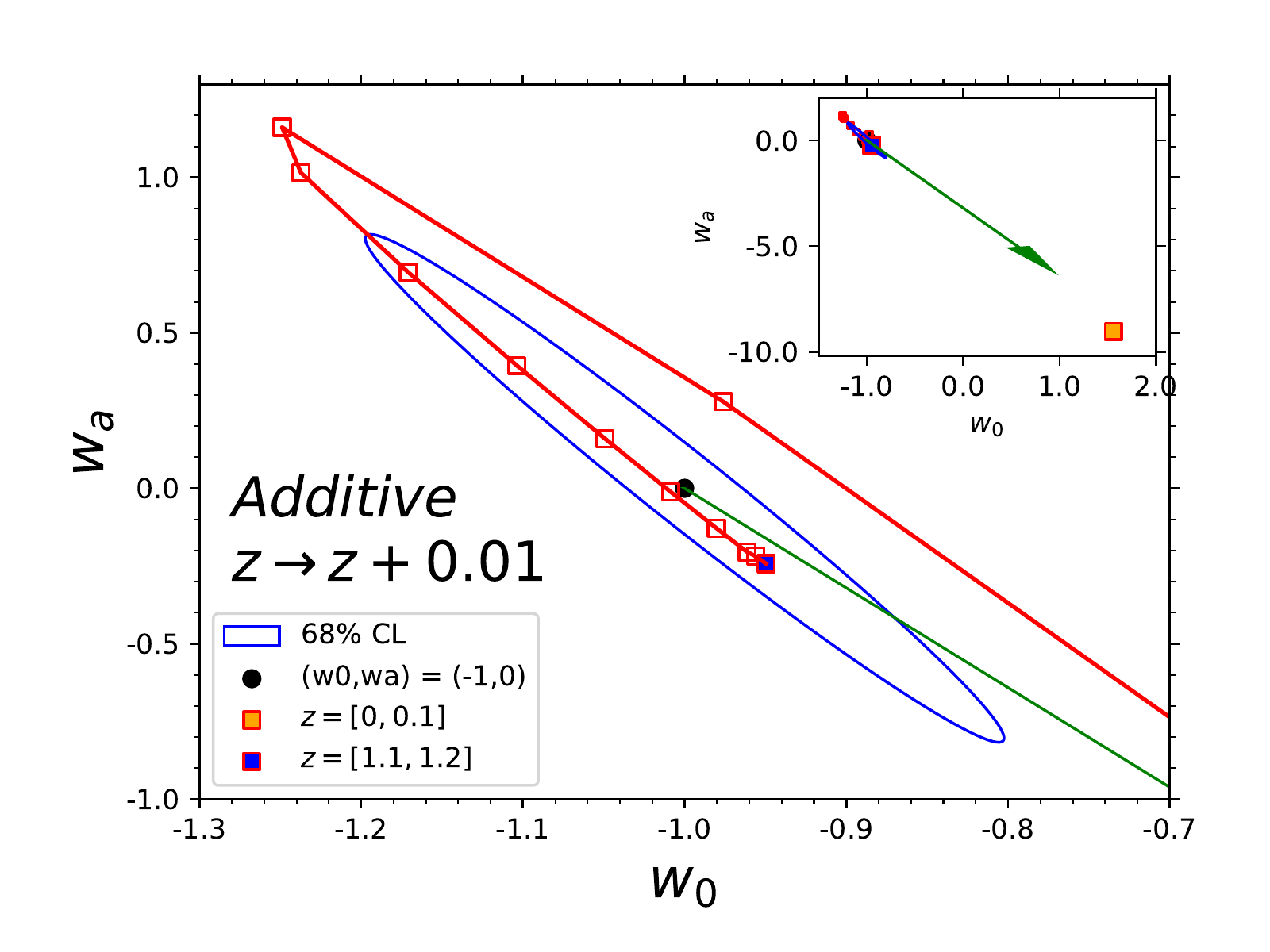}
\caption{
The dark energy parameter shifts from an additive redshift systematic, 
$(d_0,d_1)=(0.01,0)$, are 
plotted in the $w_0$--$w_a$ plane, along with the statistical $1\sigma$ 
joint confidence contour. The red curve indicates the shifts as the 
systematic is applied individually to each redshift, from lowest bin 
($z=[0,0.1]$: solid orange box) to highest ($z=[1.1,1.2]$: solid blue box), 
with open red squares every 
0.1 in redshift. Applying the systematic at all redshifts shifts the 
fiducial cosmology from the black dot ($\Lambda$CDM) to the end of the 
green arrow. We show the full extent of the shift in the inset plot. 
} 
\label{fig:addbias10}
\end{figure}

For the additive redshift systematic we take  $d_0=0.01$, $d_1=0$, i.e.\ $z\to z+0.01$ for 
illustration. We calculate the cosmological parameter bias induced 
by this systematic, applied to each of the 12 redshift bins individually, and to 
all of them. Figure~\ref{fig:addbias10} shows the results in the 
$w_0$--$w_a$ dark energy plane. Each red box shows the shift from the 
fiducial $\Lambda$CDM cosmology with $w_0=-1$, $w_a=0$ (black dot at the 
center of the blue 68.3\% joint confidence contour ellipse). The lowest 
redshift bin $z=[0,0.1]$ is marked with orange fill and the highest 
redshift bin $z=[1.1,1.2]$ is marked with blue fill; all bins are 
connected in order of redshift by the red curve.  

The largest individual bias occurs for the lowest bin but substantial 
bias is evident for several redshift bins. The green arrow gives 
the total bias for the systematic applied to all redshifts. Note the bias 
for $d_0=0.01$ is so large that it extends well beyond the $1\sigma$ 
joint confidence contour, as shown by the inset figure. We can best 
quantify the overall dark energy bias by using the $\dchi$ statistic: 
such a redshift systematic deliver $\dchi=505$, some $20\sigma$ off. 
To determine the systematic control requirement we can solve numerically 
for the condition $\dchi=2.3$ (or use that $\dchi\sim [\delta z]^2$), to 
find the requirement $d_0\lesssim 0.0006$. This is quite severe, and 
while later we will investigate ways of easing this, we see that 
photometric redshifts will be greatly challenged to give robust cosmology.

%%%%%%%%%%%%%%%%%%%%%%

%%%%%%%%%%%%%%%%%%% 
\subsection{Multiplicative Systematic} \label{sec:mult} 

\begin{figure}[htbp!] 
\centering 
\includegraphics[width=\columnwidth]{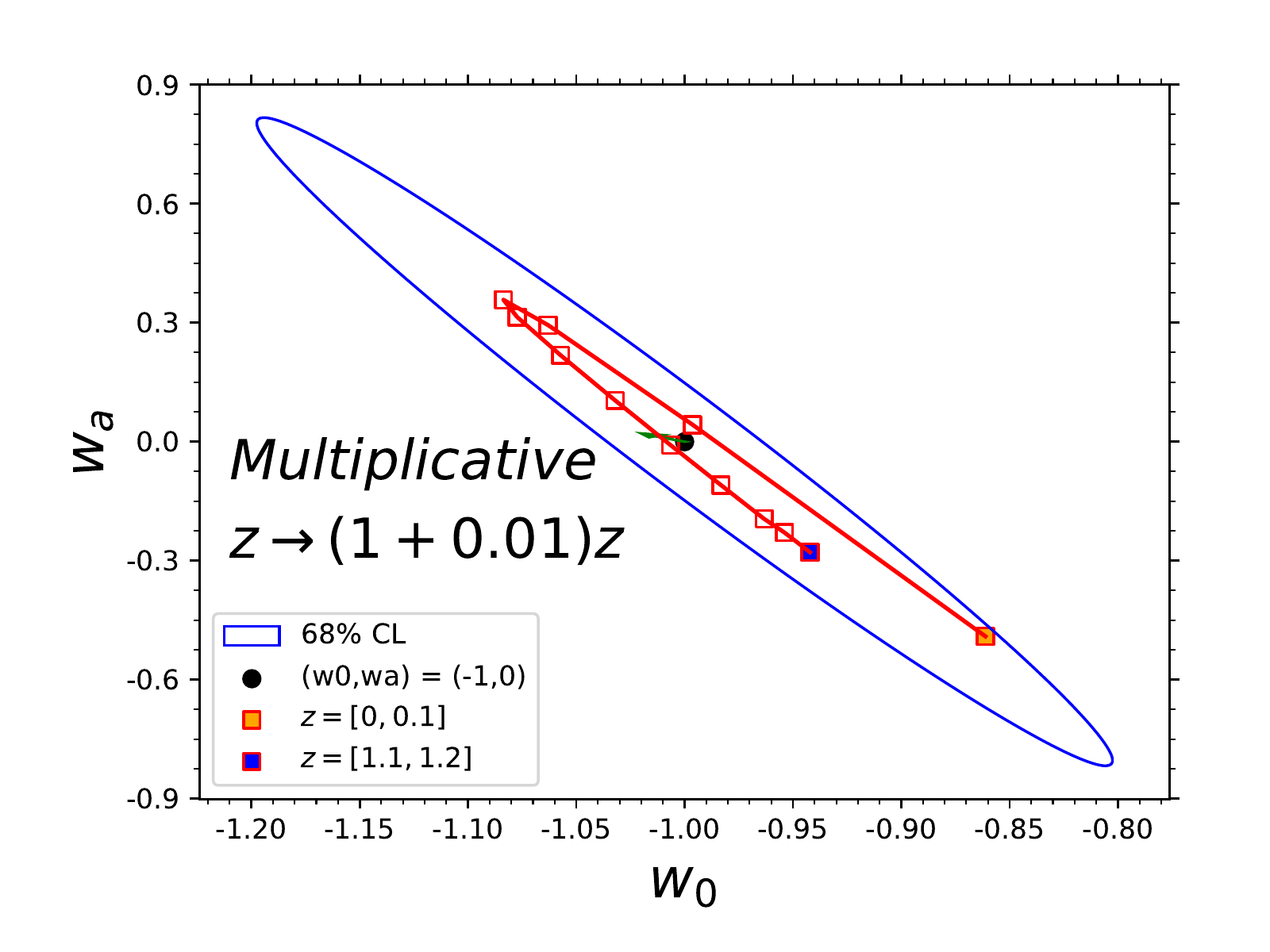} 
\caption{As Fig.~\ref{fig:addbias10} but for a multiplicative redshift 
systematic with $(d_0,d_1)$ = $(0,0.01)$. 
}
\label{fig:multbias01} 
\end{figure}

For the multiplicative redshift systematic, we take  $d_0=0$, $d_1=0.01$, i.e.\ $z\to (1+0.01)\,z$ for 
illustration. The analysis for bias induced by multiplicative systematics follows that of the previous subsection on the additive case. In Fig.~\ref{fig:multbias01} we can see that the total bias due to redshift systematic applied to all redshift 
bins is significantly smaller than the additive case. There is 
a coincidental cancellation among the (lesser) biases of individual redshift 
bins such that the sum is small, lying within the $1\sigma$ 
joint confidence contour. 

This gives a small shift $\dchi=0.4$ for $d_1=0.01$, meaning 
that $d_1\lesssim 0.024$ would satisfy the $\dchi<2.3$ 
criterion. However we emphasize that this cancellation is 
somewhat fine tuned, as we explore in the following subsection, 
and so the requirements on multiplicative systematics control 
would be better regarded as $d_1\lesssim 0.01$.

%%%%%%%%%%%%%%%%%%%% 

%%%%%%%%%%%%%%%%%%%%%%%%% 
\subsection{Systematics Control} \label{sec:control} 

Table~\ref{tab:addmul} summarizes the additive and multiplicative redshift 
systematics cases. We see the additive case is much more severe, with 
spectroscopic level requirements on the redshifts. 
Allowing for both additive and multiplicative 
systematics tightens the control requirements. (Note that a multiplicative 
systematic in $1+z$ corresponds to a combination of additive and multiplicative 
systematics in $z$.) In all cases, photometric 
precision is insufficient for robust cosmology determination. 

Since we expect a small 
redshift shift to affect low redshifts more, due to the sensitivity of 
the SN apparent magnitude to redshift at low $z$ (going roughly as $1/z$), then we 
also consider the case where spectroscopic redshifts exist for all local 
($z<0.1$) SN and so systematics there vanish. This helps substantially 
(but not enough) for the additive case, while for the multiplicative 
case it actually worsens the effect.

%%%%%%%%%%%%%%%%%%%%%%%%%%% 
\begin{table}[htbp!]
\begin{tabular}{|l|c|c|c|c|c|}
\hline 
\,Model & $\delta\om$ & $\delta w_0$ & $\delta w_a$ & \ $\dchi$ \ & $\delta z_{\rm req}$ \\ 
\hline 
\,Additive \ & 0.065 & 1.92 & $-6.14$ & 505 & \ 0.0006  \\ 
\,Additive (no local) & 0.041 & $-0.64$ & 2.90 & 33 & 0.003  \\ 
\,Mult. & \ $-0.004$ \ & \ $-0.02$ \ & 0.021 & 0.4 & 0.024 \\ 
\,Mult.~(no local) & \ $-0.005$ \ & $-0.16$ & 0.51 & 3.4 & 0.008 \\ 
\hline
 \end{tabular}
\caption{
Cosmology biases due to additive and multiplicative redshift systematics 
at the 0.01 level. The requirement on the systematic level to control bias to 
$\dchi<2.3$ ($1\sigma$ joint confidence) is given by $\delta z_{\rm req}$ 
for each case (where $\delta z_{\rm req}$ is to be interpreted as either 
$d_0$ or $d_1$). The case with no systematics for $z<0.1$ SN is shown by 
``(no local)''; note for the multiplicative case this removes a 
cancellation and tightens the requirement. 
 }
\label{tab:addmul}
\end{table}

It is worthwhile pursuing the question further as to whether systematics control can be concentrated in 
particular redshift bins. As mentioned, the lowest redshift 
bin systematic gives substantial cosmology bias in both cases. Note 
that a similar characteristic was found in \cite{0402002}. Indeed, if 
we eliminate the systematic in the $z=[0,0.1]$ bin then for the 
additive case the $\dchi$ drops from 505 to 33. Of course this is still 
far more biased than we can accept. Note however that for the 
multiplicative case, $\dchi$ actually worsens from 0.4 to 3.4, because 
the bias from the lowest bin canceled some of the bias from 
higher bins. 

\begin{figure}[htbp!] 
\centering 
\includegraphics[width=\columnwidth]{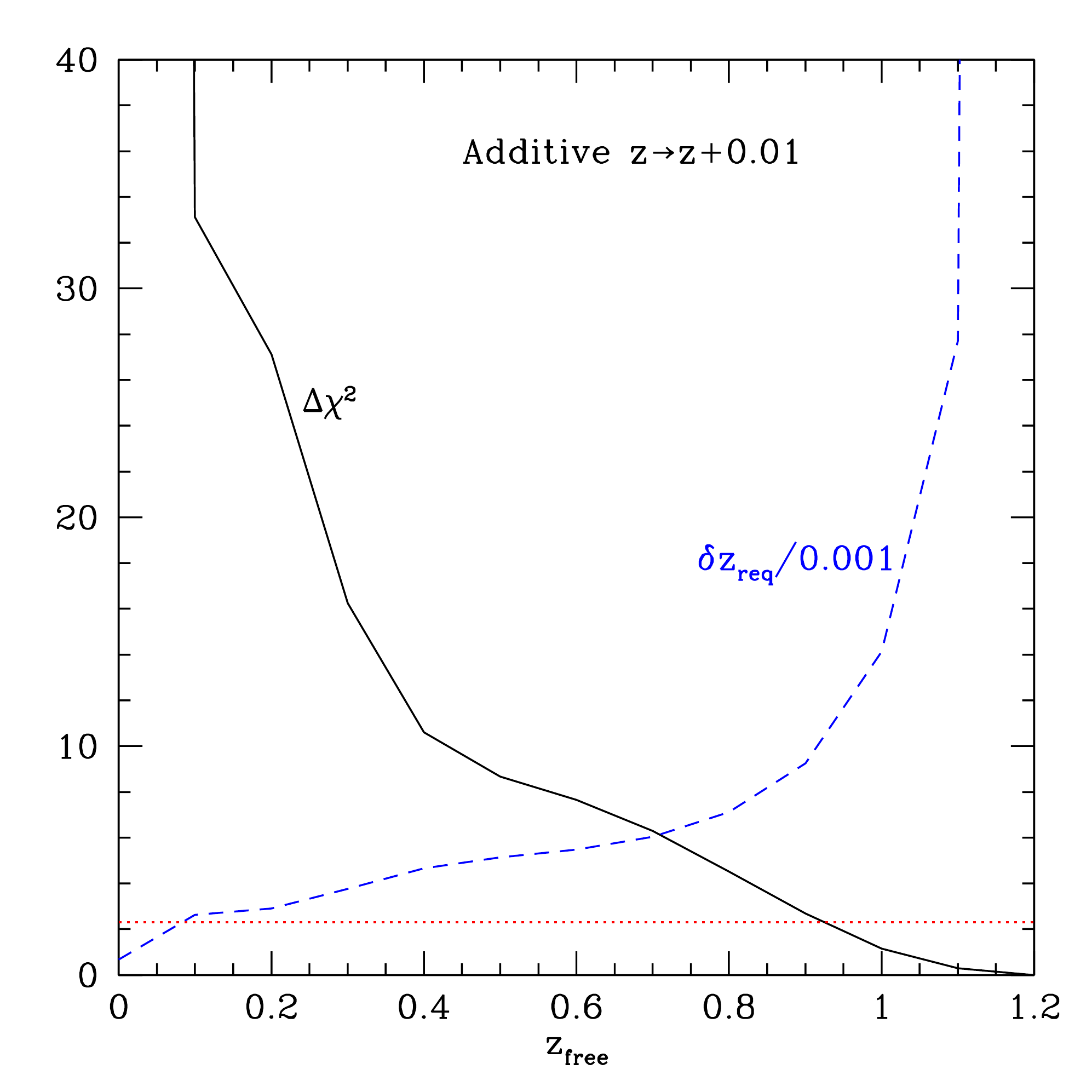} 
\caption{The relaxation of the cosmology bias $\dchi$ (solid black 
curve) and required additive redshift systematic control 
$\delta z_{\rm req}$ (dashed blue curve) is shown as a function of 
the redshift $z_{\rm free}$ out to which the systematic is eliminated, 
e.g.\ due to use of a spectroscopic sample. The dotted red line shows 
$\dchi=2.3$; to avoid substantial cosmology bias we need to work in 
the region where the solid black curve lies below the dotted red curve.
}
\label{fig:addfree} 
\end{figure}

\begin{figure}[htbp!] 
\centering 
\includegraphics[width=\columnwidth]{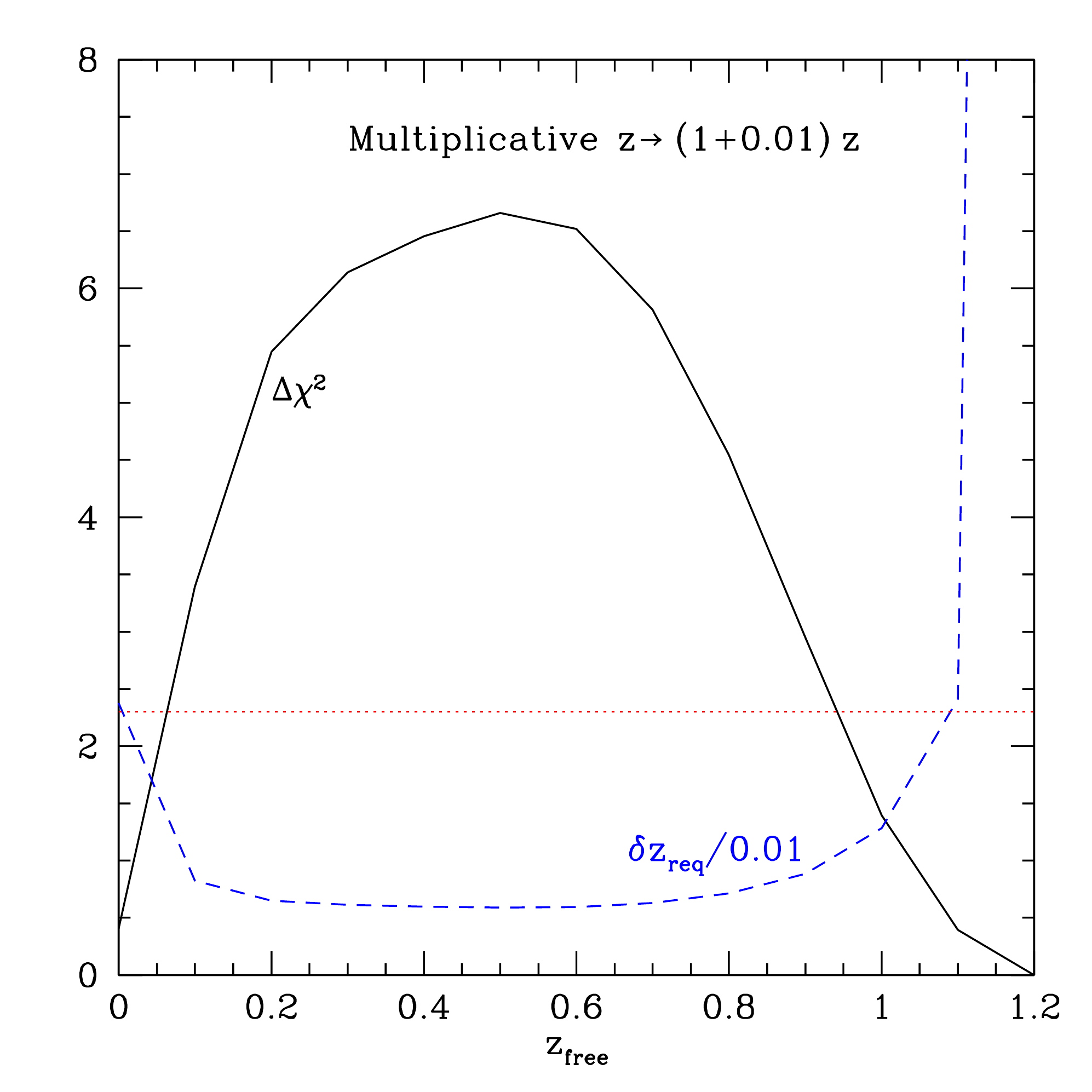} 
\caption{As Fig.~\ref{fig:addfree} but for the multiplicative 
systematic. Note the $\delta z_{\rm req}$ curve is divided by 0.01 
not 0.001 as in the additive case. 
}
\label{fig:multfree} 
\end{figure}

Figure~\ref{fig:addfree} presents the results of systematic redshift 
control, e.g.\ by use of a SN spectroscopic sample, to make all 
redshifts $z<z_{\rm free}$ 
free from additive systematics. Figure~\ref{fig:multfree} shows the 
multiplicative systematics case. We show the effect of such control 
both on $\dchi$ and on the requirement $\delta z_{\rm req}$ on the 
remaining redshifts $z>z_{\rm free}$ (compare Table~\ref{tab:addmul}). To avoid 
substantial cosmology bias we need to either eliminate systematics 
through use of a spectroscopic sample out to $z_{\rm free}\gtrsim0.9$ 
(with the higher redshift photometric sample having a systematic of 
0.01), or use of a spectroscopic sample out to $z_{\rm free}$ and a 
systematic level at higher redshifts below the $\delta z_{\rm req}$ 
curve.

%%%%%%%%%%%%%%%%%%%% 

%%%%%%%%%%%%%%%%%%%% 

Note the multiplicative case has a different behavior than the additive 
case in the shape of the $\dchi$ curve. Elimination of systematics 
for the lowest redshift bin controls cosmology bias, but this is due 
to a fine tuned cancellation. The $\dchi$ curve increases initially, 
rather than monotonically decreasing as in the additive case. So we 
cannot depend on removing systematics only from the lowest bin; if we 
removed systematics from the first two or three bins, instead the 
cosmology would be strongly biased. Only by removing systematics out 
to $z_{\rm free}\gtrsim0.9$ can we guarantee robust cosmology results. 
In all other cases we see that photometric redshift systematic control 
at the 0.01 level is insufficient; spectroscopy for the majority of 
the SN is required.

%%%%%%%%%%%%%%%%%%% 
\subsection{Catastrophic Outliers} \label{sec:cat} 

The three catastrophic outlier toy models can also give rise to a 
bias on cosmology. In Table~\ref{tab:cat} we summarize the three cases, 
showing the bias induced for a catastrophic outlier fraction of 
1\%, and also the requirement on the fraction $f$ in order to keep 
$\dchi<2.3$. Recall the values $\delta p_i$ will scale nearly linearly 
with $f$, and the $\dchi$ will scale nearly as its square.

%%%%%%%%%%%%%%%%%%%%%%%%%%% 
\begin{table}[htbp!]
\begin{tabular}{|l|c|c|c|c|c|}
\hline 
\ Model & $\delta\om$ & $\delta w_0$ & $\delta w_a$ & \ $\dchi$ \ & $f_{\rm req}$ \\ 
\hline 
\ $z\to0.1$ \ & 0.026 & 0.19 & \ $-0.36$ \ & 21 & \ 0.003 \ \\ 
\ $z\to1$ & 0.028 & 0.20 & $-0.35$ & 25 & 0.003 \\ 
\ $z\pm0.2$ & \ 0.017 \ & \ 0.099 \ & $-0.14$ & 8.1 & 0.005 \\ 
\hline
 \end{tabular}
\caption{
Cosmology biases due to various catastrophic redshift models with 1\% 
outliers. The requirement on the outlier fraction to control bias to 
$\dchi<2.3$ ($1\sigma$ joint confidence) is given by $f_{\rm req}$ for 
each case. 
 }
\label{tab:cat}
\end{table}

\begin{figure}[htbp!]
\includegraphics[width=\columnwidth]{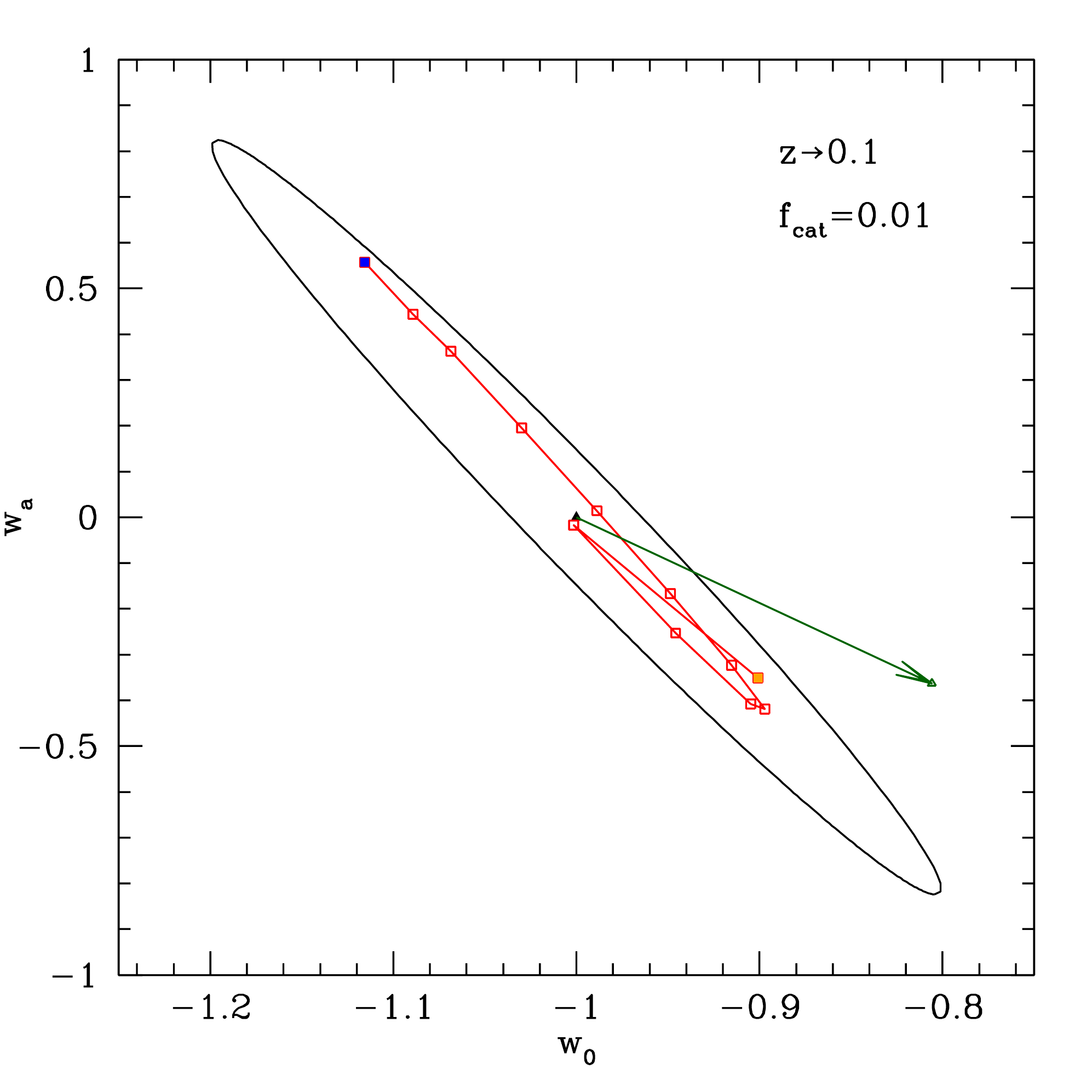} 
\caption{
As Fig.~\ref{fig:addbias10} but for a catastrophic redshift 
systematic with outlier fraction 
$f_{\rm cat}=0.01$ misinterpreted as being at $z=0.1$. 
} 
\label{fig:catarrow01}
\end{figure}

\begin{figure}[htbp!]
\includegraphics[width=\columnwidth]{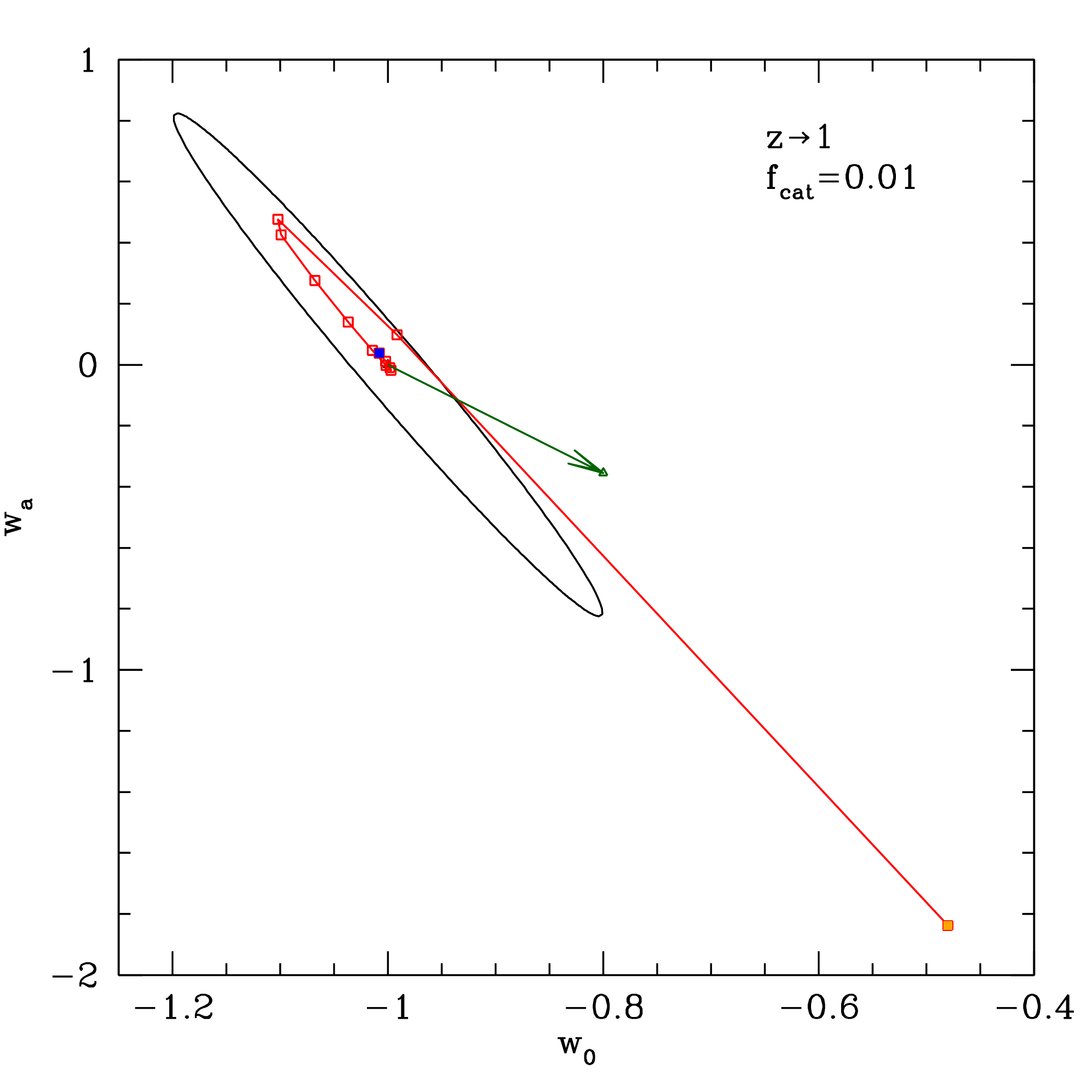} 
\caption{As Fig.~\ref{fig:addbias10} but for a catastrophic redshift 
systematic with outlier fraction $f_{\rm cat}=0.01$ misinterpreted as 
being at $z=1$. Note the huge shift due to the lowest redshift bin 
outliers. 
} 
\label{fig:catarrow10}
\end{figure}

These results hold for catastrophic redshift outliers in every bin. 
However, it is useful to break this down to investigate the contribution 
from each individual 
redshift bin. Figures~\ref{fig:catarrow01} and \ref{fig:catarrow10} 
show an interesting effect. For the $z\to0.1$ case, although all the 
redshift outliers at each redshift bin give small shifts lying within 
the 68\% joint confidence contour (i.e.\ $\dchi<2.3$), more of them lie 
to the lower right so the summed cosmological parameter shift amounts to 
$\dchi=21$, or approximately $4.5\sigma$. (Recall that the $\dchi$ do 
not sum linearly.) For the $z\to1$ case, all but the lowest redshift bin 
do not give large parameter shifts. However the lowest redshift bin gives 
a very strong dark energy shift (which to some extent is canceled by 
the shift in the opposite direction by the other redshift bins). 
If the first bin were systematics free, then $\dchi$ drops from 24.6 to 
9.5, meaning the requirement on $f$ for the other bins loosens to 0.005. 
(For the $z\to 0.1$ case the lowest $z$ bin systematic does not give a 
strong effect.)

%%%%%%%%%%%%%%%%%%%%%%%% 

%%%%%%%%%%%%%%%%%%%%%%%% 

\begin{figure}[htbp!]
\includegraphics[width=\columnwidth]{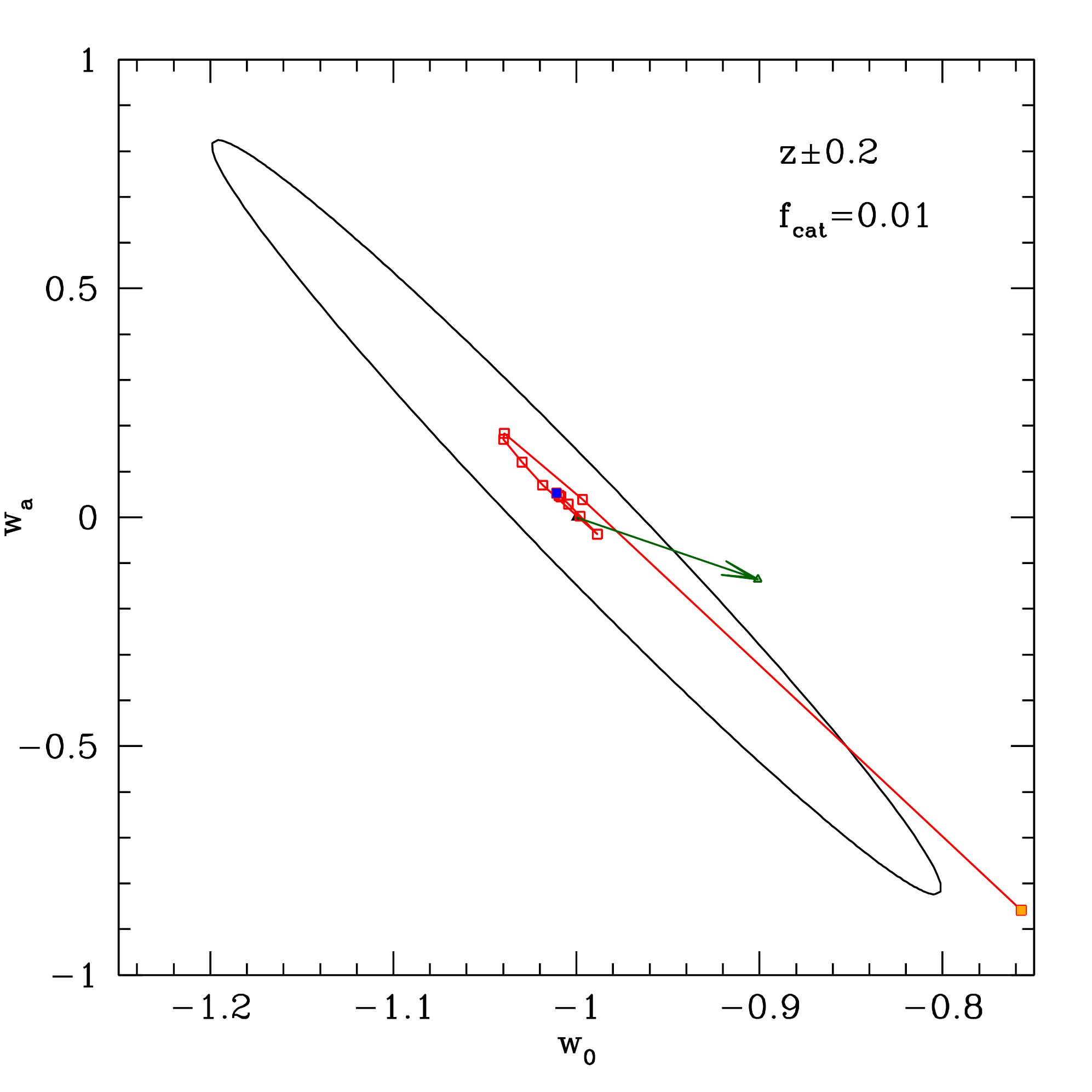} 
\caption{As Fig.~\ref{fig:addbias10} but for a catastrophic redshift 
systematic with outlier fraction $f_{\rm cat}=0.01$ misinterpreted as 
being at $z+0.2$ for $z\le0.6$ and $z-0.2$ for $z>0.6$, i.e.\ losing 
from the extremes. Note the large shift due to the lowest redshift bin 
outliers. 
} 
\label{fig:catarrowpm2}
\end{figure}

For the third model, with $z\to z\pm0.2$, again we find that catastrophic 
outliers in the lowest redshift bin are the most damaging, and again it 
is partially controlled by opposite shifts from the other bins, as seen 
in Fig.~\ref{fig:catarrowpm2}. If the 
first bin were systematics free, then $\dchi$ drops from 8.1 to 3.2 
(and $f_{\rm req}$ rises to 0.008).

%%%%%%%%%%%%%%%%%%%%%%%% 

%%%%%%%%%%%%%%%%%%%%%%%%%%%%%%%%%%%%%%%%%% 
\section{Conclusion} \label{sec:concl} 

Tens to hundreds of thousands of Type Ia supernovae will be discovered by 2020s 
wide area imaging surveys such as ZTF and LSST. While these provide some information 
on the SN, they do not provide spectral information which is useful for 
classification and subclassification, and redshifts. Follow up spectroscopy, even for 
host galaxy redshifts, is expensive in terms  of telescope time. We investigate specifically 
the issue of cosmology bias due to systematically imperfect redshift determination, 
quantifying the cosmology bias and resulting requirements for systematic control. 

We examine three classes of redshift systematics -- additive, multiplicative, and 
catastrophic -- and conclude that in all cases robust cosmology requires control 
of redshift systematics at the subpercent level. We show how cosmology bias evolves 
as the systematic enters at different redshifts, and generally the sum over the full 
sample leads to large offsets from the true cosmology. Investigating whether limited 
spectroscopy in the form of a focus on particular redshift bins, e.g.\ low redshift, 
removes the issue gives the conclusion that it generally does not; systematic 
control throughout the sample is essential. This analysis is complementary to that 
of \cite{0402002}, which examined the ``bloat'' of uncertainties rather than bias, 
and came to similar accuracy requirement conclusions. 

Additive systematics appear the most harmful, and then certain types of catastrophic 
outliers. Multiplicative redshift systematics initially appear relatively benign, but this is 
due to a fine tuned cancellation and small deviations from the model do impose  
subpercent control requirements. A spectroscopic sample, though more limited in 
numbers, is needed for robust cosmology determination. Many spectroscopic 
instruments will be operating during the 2020s, such as the Dark Energy Spectroscopic 
Instrument (DESI) and Wide Field Infrared Survey Telescope (WFIRST), and could play 
useful roles in contributing to a well controlled sample of several thousand SN, 
capable of high accuracy constraints on dark energy and cosmology.

%%%%%%%%%%%%%%%%%%%%%%%%%%%%%% 
\section*{Acknowledgments}

We thank Julien Guy and Alex Kim for helpful conversations. 
This work is supported in part by the Energetic Cosmos Laboratory and by 
the U.S.\ Department of Energy, Office of Science, Office of High Energy 
Physics, under Award DE-SC-0007867 and contract no.\ DE-AC02-05CH11231.


\begin{thebibliography}{99}

\bibitem{perl99} 
S. Perlmutter et al., ApJ 517, 565 (1999) [arXiv:astro-ph/9812133] 

\bibitem{riess98} 
A.G. Riess et al., Astron. J. 116, 1009 (998) [arXiv:astro-ph/9805201] 

%\bibitem{schmidt} 
%Schmidt, Brian P., et al. ApJ 507.1 (1998): 46. 

%\bibitem{r0}
%Astier, Pierre, et al. Astron. Astrophys. 447.1 (2006): 31-48.
%\bibitem{r1}
%Kowalski, Marek, et al., ApJ 686.2 (2008): 749.

%\bibitem{r2}
%Jones, D. O., et al., ApJ 857.1 (2018): 51.

\bibitem{review1} 
D. Huterer, D.L. Shafer, Rep. Prog. Phys. 81, 016901 (2018) [arXiv:1709.01091] 

\bibitem{review2} 
B. Leibundgut, M. Sullivan, Space Sci. Rev. 214, 57 (2018)  [arXiv:astro-ph/0003326] 

\bibitem{union2p1} 
N. Suzuki et al., ApJ 746, 85 (2012) [arXiv:1105.3470] 

%Amanullah, Rahman, et al., ApJ 716.1 (2010): 712. [arXiv:astro-ph/1004.1711]

\bibitem{jla} 
M. Betoule et al., Astron. Astrophys. 568, A22  (2014) [arXiv:1401.4064] 

\bibitem{pantheon} 
D.M. Scolnic et al., ApJ 859, 101 (2018) [arXiv:1710.00845] 

%\bibitem{wood}
%Wood-Vasey, W. Michael, et al. ApJ 666.2 (2007): 694.

%\bibitem{scol}
%Scolnic, D., et al. ApJ 815.2 (2015): 117.

\bibitem{ztf} 
E.C. Bellm et al., PASP 131, 018002 (2018) [arXiv:1902.01932]  

\bibitem{lsst} 
LSST Dark Energy Science Collaboration, [arXiv:1809.01669] 
%Abell, Paul A., et al., SLAC-R-1031.(2009) [arXiv:0912.0201] 

\bibitem{snir} 
E.Y. Hsiao et al., PASP 131, 014002 (2018) [arXiv:1810.08213] 

\bibitem{snir1} 
A. Avelino, A.S. Friedman, K.S. Mandel, D.O. Jones, P.J. Challis, R.P. Kirshner, arXiv:1902.03261 

%\bibitem{snir1}
%Maguire, K., et al.  MNRAS 477.3 (2018): 3567-3582.

%\bibitem{snir2} 
%Roberts-Pierel, Justin, Steven A. Rodney, and Steven Rodney.  BAAS 231. Vol. 231. 2018.  

\bibitem{snuv} 
E.S. Walker, S. Hachinger, P.A. Mazzali, R.S. Ellis, M. Sullivan, A. Gal-Yam, D.A. Howell, MNRAS 427, 103 (2012) [arXiv:1208.4130]

%\bibitem{snuv1} 
%Maguire, K., et al. MNRAS 426.3 (2012): 2359-2379.

\bibitem{snuv1} 
R.J. Foley et al., MNRAS 461, 1308 (2016) [arXiv:1604.01021] 

%\bibitem{snspec0}
%Branch, David, and G. A. Tammann.  Annu. Rev. Astron. Astrophys. 30.1 (1992): 359-389.

%\bibitem{snspec1} 
%Walker, E. S., et al. MNRAS 410.2 (2010): 1262-1282.

%\bibitem{snspec2}
%Walker, E. S., et al. MNRAS 410.2 (2010): 1262-1282.

%\bibitem{snspec3}
%Hoeflich, Peter, J. C. Wheeler, and F. K. Thielemann. ApJ 495.2 (1998): 617.

%\bibitem{snspec4}
%Phillips, M. M., et al. ASP 99.617 (1987): 592.

\bibitem{snspec} 
D. Brout et al., ApJ 874, 150 (2019) [arXiv:1811.02377] 

\bibitem{snspec1} 
M. Sasdelli et al., MNRAS 447, 1247 (2015) [arXiv:1411.4424] 

\bibitem{snspec2} 
C. Saunders et al., ApJ 869, 167 (2018) [arXiv:1810.0947] 

%\bibitem{snspec5}
%Sako, Masao, et al. ASP 130.988 (2018): 064002.

%\bibitem{snspec6}
%T. J. Bronder et al. Astron. Astrophys. 477.3 (2008): 717-734.[arXiv:0709.0859v1]

%\bibitem{saulpt} 
%Perlmutter and Schmidt Physics today 

\bibitem{likes} 
D. Branch, S. Perlmutter, E. Baron, P. Nugent,  arXiv:astro-ph/0109070 

\bibitem{likes2} 
G. Garavini et al., Astron. Astrophys. 470, 411 (2007) [arXiv:astro-ph/0703629] 

\bibitem{0812.0370} 
E.V. Linder, Phys. Rev. D 79, 023509 (2009) [arXiv:0812.0370] 

\bibitem{hannah} 
H.K. Fakhouri, ApJ 815, 58 (2015) [arXiv:1511.01102] 

%Fakhouri, Hannah. Diss. UC Berkeley, (2013). 

\bibitem{hannah2} 
D. Rubin, arXiv:1903.10518 

%Boone, Kyle, et al. AAS 227. Vol. 227. (2016). 

\bibitem{des} 
E. Macaulay et al., MNRAS 486, 2184 (2019) [arXiv:1811.02376] 

%\bibitem{desi} 
% Brenna Flaugher and Chris Bebek, SPIE V. Vol. 9147. (2014).[arXiv:1812.00515]

\bibitem{desi}
DESI Collaboration, arXiv:1611.00036 

%\bibitem{desi2} 
%Mariana Vargas-Magana et al. (2019) [arXiv:1901.01581].

\bibitem{desi2}
DESI Collaboration, arXiv:1611.00037 

\bibitem{4most} 
E. Swann et al., Messenger 175, 68 (2019) [arXiv:1903.02476] 

%\bibitem{syst0}
%Alex G. Kim et al. MNRAS 347.3 (2004): 909-920.[arXiv:astro-ph/0304509]

%\bibitem{syst1} 
%Bonnie R. Zhang et al. MNRAS 471.2 (2017): 2254-2285.[arXiv:1706.07573]

%\bibitem{syst2}
%Adam G. Riess et al. ApJ 730.2 (2011): 119.[arXiv:1103.2976]

%\bibitem{syst3}
%George Efstathiou. MNRAS 440.2 (2014): 1138-1152.[arXiv:1311.3461]

\bibitem{syst1} 
A. M{\"o}ller et al., JCAP 1612, 8 (2016) [arXiv:1608.05423] 

\bibitem{syst2} 
M. Dai, S. Kuhlmann, Y. Wang, E. Kovacs, MNRAS 477, 4142 (2018) [arXiv;1701.05689] 

\bibitem{syst3} 
E. Roberts, M. Lochner, J. Fonseca, B.A. Bassett, P-Y. Lablanche, S. Agarwal, JCAP 1710, 36 (2017) [arXiv:1704.07830] 

\bibitem{syst4} 
H.S. Xavier et al., MNRAS 444, 2313 (2014) [arXiv:1312.5706]

\bibitem{syst5} 
M. Sako et al., ApJ 738, 162 (2011) [arXiv:1107.5106] 

\bibitem{syst6} 
R. Kessler et al., arXiv:1903.11756 

\bibitem{syst7} 
V.A. Villar et al., arXiv:1905.07422 

\bibitem{syst7b} 
J. Lasker et al., MNRAS 485, 5329 (2019) [arXiv:1811.02380] 

%\bibitem{syst5}
%J. Guy et al., Astron. Astrophys. 523, 7 (2010) [arXiv:1010.4743]

%\bibitem{syst6}
%tDaniel Scolnic et al. ApJ 795.1 (2014): 45.[arXiv:1310.3824]

%\bibitem{syst7}
%A. Conley  et al.  Astrophys. J., Suppl. Ser. 192.1 (2010): 1.

\bibitem{syst8}
N. Palanque-Delabrouille et al., Astron. Astrophys. 514, 63 (2010) [arXiv:0911.1629] 

\bibitem{syst9}
D.O. Jones et al., ApJ 843, 6 (2017) [arXiv:1611.07042] 

\bibitem{syst10}
D.O. Jones et al., ApJ 857, 51 (2018) [arXiv:1710.00846] 

\bibitem{syst11}
D.O. Jones et al., arXiv:1811.09286 

\bibitem{hosts} 
R. Gupta et al., Astron. J. 152, 154 (2016) [arXiv:1604.06138] 

\bibitem{hosts1} 
S.A. Uddin, J. Mould, C. Lidman, V. Ruhlmann-Kleider, D. Hardin, PASA 34, 9 (2017) [arXiv:1612.07883] 

\bibitem{hosts2} 
M. Roman et al., Astron. Astrophys. 615, A68 (2018)  [arXiv:1706.07697] 

%\bibitem{hosts2}
%Mannucci, Filippo, et al. Astron. Astrophys. 433.3 (2005): 807-814.

%\bibitem{hosts3}
%Sullivan, Mark, et al. ApJ 648.2 (2006): 868.

%\bibitem{hosts4}
%Neill, James D., et al. ApJ 707.2 (2009): 1449.

%\bibitem{hosts5}
%Howell, D. Andrew, et al. ApJ 691.1 (2009): 661.

%\bibitem{hosts6}
%Gallagher, Joseph S., et al. ApJ 634.1 (2005): 210.

%\bibitem{hosts7}
%Hamuy, Mario, et al. (1996) [astro-ph/9609063].

%\bibitem{hosts8}
%Hamuy, Mario, et al. (1996) [astro-ph/9609062].

%\bibitem{hosts9}
%Pan, Y-C., et al. MNRAS 446.1 (2014): 354-368.

\bibitem{0402002} 
D. Huterer, A. Kim. L.M. Krauss, T. Broderick, ApJ 615, 595 (2004) [arXiv:astro-ph/0402002] 

\bibitem{0607030} 
G. Aldering, A.G. Kim, M. Kowalski, E.V. Linder, S. Perlmutter, Astropart. Phys. 27, 213 (2007) [arXiv:astro-ph/0607030] 

\bibitem{fisbias1} 
L. Knox, R. Scoccimarro, S. Dodelson, Phys. Rev. Lett. 81, 2004 (1998) [arXiv:astro-ph/9805012] 

\bibitem{fisbias2} 
E.V. Linder, Astropart. Phys. 26, 102 (2006) [arXiv:astro-ph/0604280]  

\bibitem{0908.2637} 
J. Samsing, E.V. Linder, Phys. Rev. D 81,  043533 (2010) [arXiv:0908.2637] 

\bibitem{kcrx1} 
T.M. Davis, B.P. Schmidt, A.G. Kim, PASP 118, 205 (2006) [arXiv:astro-ph/0511017] 

\bibitem{kcrx2} 
E.Y. Hsiao et al., ApJ 663, L13 (2007) [arXiv:astro-ph/0703529] 

\bibitem{salt2} 
J. Guy et al., Astron. Astrophys. 466, 11 (2007) [arXiv:astro-ph/0701828] 

\bibitem{shapiro} 
S. Dodelson, C. Shapiro, M. White, Phys. Rev. D 73, 023009 (2006) [arXiv:astro-ph/0508296] 

\bibitem{shapiro2} 
C. Shapiro, ApJ 696, 775 (2009) [arXiv:0812.0769] 



\end{thebibliography}
\end{document}